\documentstyle[12pt]{article}
\def\gsim{\mathrel{\rlap {\raise.5ex\hbox{$ > $}}
{\lower.5ex\hbox{$\sim$}}}}
\def\lsim{\mathrel{\rlap {\raise.5ex\hbox{$ < $}}
{\lower.5ex\hbox{$\sim$}}}}

\newcommand{\be}{\begin{equation}}
\newcommand{\ee}{\end{equation}}
\newcommand{\bea}{\begin{eqnarray}}
\newcommand{\nn}{\nonumber}
\newcommand{\eea}{\end{eqnarray}}

\def\gappeq{\mathrel{\rlap {\raise.5ex\hbox{$>$}}
{\lower.5ex\hbox{$\sim$}}}}
 
\def\lappeq{\mathrel{\rlap{\raise.5ex\hbox{$<$}}
{\lower.5ex\hbox{$\sim$}}}}
 
\begin{document}
 
\begin{titlepage}
\begin{flushright}
ACT-5/98 \\
CTP-TAMU-14/98 \\
OUTP--98--27P \\
hep-th/9804084 \\
\end{flushright}

\begin{centering}
\vspace{.1in}
{\large {\bf World-Sheet Formulation of $M$ Theory }} \\
\vspace{.2in}

{\bf John Ellis$^{a}$},  
{\bf N.E. Mavromatos$^{b}$} 
and 
{\bf D.V. Nanopoulos$^{c}$}

\vspace{.03in}
 
{\bf Abstract} \\
\vspace{.1in}
\end{centering}
{\small 
We first review the interpretation of world-sheet defects as $D$ branes
described by a critical theory in 11 dimensions, that we interpret as
$M$ theory. We then show that $D$-brane recoil induces
dynamically an anti-de-Sitter (AdS) space-time background, with criticality
restored by a twelfth time-like dimension described by a Liouville field.
Local physics in the bulk of this AdS$_{11}$ may be described by an
$Osp(1|32,R) \otimes Osp(1|32,R)$ topological gauge theory (TGT), with
non-local boundary states in doubleton representations. We draw analogies
with structures previously exhibited in two-dimensional black-hole
models. Wilson loops of `matter' in the TGT may be described by an effective
string action, and defect condensation may
yield string tension and cause a space-time metric to appear.}

\vspace{0.15in}
\begin{flushleft}
$^{a}$ Theory Division, CERN, 1211 Geneva 23, Switzerland. \\
$^{b}$ P.P.A.R.C. Advanced Fellow, Department of Physics
(Theoretical Physics), University of Oxford, 1 Keble Road,
Oxford OX1 3NP, U.K.  \\
$^{c}$ Department of Physics, 
Texas A \& M University, College Station, \\
TX~77843-4242, USA, \\
Astroparticle Physics Group, Houston
Advanced Research Center (HARC), Mitchell Campus,
Woodlands, TX 77381, USA, and \\
Academy of Athens, Chair of Theoretical Physics, 
Division of Natural Sciences, 28~Panepistimiou Avenue, 
Athens 10679, Greece. \\

\vspace{0.2in}

April 1998

\end{flushleft}

\end{titlepage} 

\newpage

\section{Introduction}

The recent exciting developments in the understanding of
non-perturbative effects in the theory formerly known as strings~\cite{dbranes}
have led to tantalizing glimpses of a broader framework
for the Theory of Everything (TOE),
referred to variously as $M$ or $F$ theory. This TOE has been
approached from several different perspectives. First were the
strong-coupling limits of various string theories that had been
thought distinct before the advent of duality. Consideration of
Type IIA string led to $M$ theory~\cite{duff} 
and that of Type IIB string led to
$F$ theory. A second perspective has been provided by the low-energy
limit. In the case of $M$ theory, 
which is Lorentz invariant for
non-trivial reasons, this low-energy limit leads
to 11-dimensional $N=1$ supergravity. The
low-energy limit of $F$ theory is less evident: it is not Lorentz
invariant, and the relation to a candidate higher-dimensional
supergravity theory is not yet clear. An interesting third
perspective has been provided by Matrix theory~\cite{matrix}, 
which proposes a
non-perturbative formulation of $M$ theory using light-cone quantization.

We have offered a fourth perspective~\cite{emndbmonop}, based on the world-sheet
$\sigma$-model formulation of string theory. By extending the
space of conformal field theories describing critical string
theories to general renormalizable two-dimensional field theories,
one is able to address issues in non-critical string theory~\cite{aben,ddk}. 
The
world-sheet renormalization scale may be identified 
with a
Liouville field~\cite{emn} whose dynamics is non-trivial 
away from
criticality. One example of
an interesting world-sheet field theory is the non-compact
Wess-Zumino model~\cite{wittenbh} 
that describes a black hole in $1+1$ dimensions,
which may be formulated as a monopole defect on the world 
sheet~\cite{emnmonop}.
We have recently shown that the supersymmetrization of this model
is conformal in 11 dimensions~\cite{emndbmonop}, 
which we interpret as the world-sheet
description of the massless solitons that appear in the strong-coupling
string approach to $M$ theory. This analysis is reviewed in more detail
below, together with some indications that a twelfth
dimension might be described by a Liouville field.

A fifth perspective on $M$ theory has been the recent suggestion~\cite{horava} 
that it might be equivalent at short distances to a Chern-Simons
topological gauge theory (TGT) based on the supergroup $Osp(1|32,R)  \otimes
Osp(1|32,R) $. The idea that a TGT might underly quantum gravity has
been recurrent in recent years. One such theory was found at the core
of the $(1+1)$-dimensional string black-hole model~\cite{eguchi,emnorigin}, 
and it provides a
natural incarnation of the holographic principle. The provocative
proposal of Horava~\cite{horava} raises several questions: is the choice of
supergroup unique? what other degrees of freedom are present in $M$
theory? what is the nature of the non-trivial dynamics that leads to the
generation of space-time? and many others.

The purpose of this paper is to address these questions from the
world-sheet perspective outlined above, building a bridge to the TGT
perspective that also illuminates many aspects of the relationship
between $M$ theory and $F$ theory. We recall that the 
appropriately supersymmetrized world-sheet
monopole corresponds to a target-space $D$ brane, and point out that
its recoil induces an anti-de-Sitter (AdS) metric in the 11-dimensional
target space, whose corresponding supergroup must be at least as
large as $Osp(1|32,R) $. Within this approach, there is an extra
time-like dimension parametrized by a Liouville field.
We point out that~\cite{holten}
the minimal supergroup
extension of the Lorentz group in $11+1$ 
dimensions is $Osp(1|32,R)  \otimes Osp(1|32,R) $, which is broken to
$Osp(1|32,R) $ because of the Lorentz non-invariance of $F$ theory. 
Horava's TGT is a local short-distance field theory with the
$Osp(1|32,R)  \otimes Osp(1|32,R) $ symmetry, but the full $M$-theory
dynamics involves non-local structures that may be expressed as
Wilson loops on the boundary of the 11-dimensional adS space.
Singleton~\cite{singl,guna,guna2} 
and higher infinite-dimensional unitary representations
of $Osp(1|32,R) $
describe non-local
boundary states. We recall that supersymmetric Wilson loops have a
string interpretation~\cite{awada}, 
in terms of which the world-sheet
monopoles characterize defects at the interface with AdS space.

The layout of this paper is as follows. In section 2, we review
aspects of our previous analysis of the two-dimensional string
black-hole model that later find echos in our analysis of $M$ theory.
In section 3 we discuss the critical monopole and vortex 
deformations in
superstring in 11 dimensions, 
and use $D$-brane recoil to derive 
in section 4 the corresponding AdS$_{11}$ metric on
target space-time. In section 5 we motivate the 
appearance of the $Osp(1|32,R)  \otimes Osp(1|32,R) $
supergroup structure in the short-distance TGT, and in section 6
we draw parallels between AdS black holes and our earlier 
two-dimensional work. In section 7 we develop the interpretation of
strings as Wilson loops in this TGT, and in 
section 8 we make some conjectures and advertize open issues within the
perspective developed here.

\section{Black Holes as World-Sheet Monopole Defects, and the
Appearance of TGT at the Core}

A relevant precursor of the discussion in this
paper is found in the two-dimensional black-hole model. The
Euclidean target-space version 
may be described in terms of a vortex defect on the world 
sheet~\cite{sathiap,ovrut,emnmonop},
obtained as the solution $X_v$ of the equation
\begin{equation}
\partial_z {\bar \partial}_z X_v = {i \pi q_v \over 2} [ \delta(z - z_1) -
\delta(z - z_2)]
\label{defect}
\end{equation}
where $q_v$ is the vortex charge and $z_{1,2}$ are the locations of
a vortex and antivortex, respectively, which we may map to the
origin and the point at infinity. The corresponding
solution to (\ref{defect}) is
\begin{equation}
X_v = q_v {\rm Im~ln} z
\label{solution}
\end{equation}
and we see that the vortex charge $q_v$ must be integer. To
see that the solution (\ref{solution}) corresponds to a black hole,
we introduce the space-time coordinates $(r,\theta)$:
\begin{equation}
z \equiv (e^r - e^{-r})e^{i\theta}
\label{transform}
\end{equation}
in terms of which the induced target-space metric is
\begin{equation}
d s^2 = {dz d{\bar z} \over 1 + z {\bar z}} = dr^2 + {\rm tanh}^2 r d
\theta^2
\end{equation}
which we recognize as a Euclidean black hole located at $r = 0$.
We recall that this model can be regarded as an $SL(2,R) / U(1)$
Wess-Zumino coset model, with gauge field
\begin{equation}
A_z \rightarrow \epsilon^2 \partial_z \theta
\label{monopole}
\end{equation}
as $r \equiv \epsilon \rightarrow 0$. We see that the
gauge field is singular at the origin, so that the world-sheet defect
may be interpreted as a monopole of the compact $U(1)$ gauge group.

There are related `spike' configurations which are solutions of the
equation
\begin{equation}
\partial_z {\bar \partial}_z X_m = - {\pi q_m \over 2} [ \delta(z - z_1) -
\delta(z - z_2)]
\label{spike}
\end{equation}
given by
\begin{equation}
X_m = q_m {\rm Re~ln} z
\label{solution2}
\end{equation}
It is easy to see that single-valuedness of the partition function imposes
the following quantization condition:
\begin{equation}
2 \pi \beta q_v q_m = {\rm integer}
\label{quantization}
\end{equation}
at finite temperature $T \equiv \beta^{-1} \ne 0$.
Making the change of variables
\begin{equation}
|z|^2 \equiv - u v \;\;: \;\; u = e^{R+t}, v = - e^{R-t}
\label{uandv}
\end{equation}
and identifying $R \equiv r + {\rm ln}(1 - e^{-r})$, we
see that the `spike' (\ref{solution2}) corresponds to a
Minkowski black hole~\cite{emnmonop}:
\begin{equation}
ds^2 = { dz d{\bar z} \over 1 + z {\bar z}} = 
 - {du dv \over 1 - uv} = dr^2 - {\rm tanh}^2 r dt^2
\label{Minkowski}
\end{equation}
The coordinates $u,v$ are natural in the Wess-Zumino
description of the Minkowski black hole, which involves
gauging the non-compact $O(1,1)$ subgroup of $SL(2,R)$.
Reparametrizing the neighbourhood of the singularity by
$w \sim {\rm ln}u \sim - {\rm ln}v$, one finds a
topological gauge theory (TGT) on the world sheet:
\begin{equation}
S_{CS} = i \int d^2z \sqrt{h} {k \over 2 \pi} w \epsilon^{ij} F(A)_{ij} +
\dots
\label{chernsimons}
\end{equation}
where $h$ is the world-sheet metric, $F$ is the field strength
of the non-compact Abelian gauge field, and the dots represent
additional `matter' or `magnon' fields. Their generic form close to the
singularity is~\cite{wittenbh}
\begin{equation}
- {k \over 2 \pi} \int d^2z \sqrt{h} h^{ij} D_i a D_j b + \dots
\label{matter}
\end{equation}
where $D_i$ is an $O(1,1)$ covariant derivative and $ab + uv = 1$.
We see from this that a non-zero condensate $<ab> \ne 0$
corresponds to $<uv> \ne 1$ and hence a non-trivial target space-time
metric (\ref{Minkowski}).

In this connection, we recall
that there appears an enhanced symmetry
at the singularity of the black hole~\cite{emnorigin}, since
the topological world-sheet theory describing the singularity 
is characterized by a target $W_{1+\infty} \otimes W_{1+\infty}$ 
symmetry. When the space-time metric is generated 
away from the singularity, there is a spontaneous breaking of
$W_{1+\infty} \otimes W_{1+\infty} \rightarrow
W_{1+\infty}$ due to the expectation value $<ab> \ne 0$,
associated with Wilson loops surrounding the world-sheet defects. 

This is a convenient point to preview the two key ways in
which the two-dimensional black hole model 
described above is relevant to the
construction of $M$ theory. The world sheet may be mapped onto
the target space-time in two dimensions, and this
two-dimensional example
may usefully be viewed from either perspective.
On the world sheet, as we discuss in the next section, 
when the above monopole solution is supersymmetrized,
it becomes a marginal deformation when the string is
embedded in an 11-dimensional space-time. 
We have suggested previously that this limit
corresponds to the masslessness of the $D$-brane representation
of target-space black holes in the strong-coupling limit of
$M$ theory when it becomes 11-dimensional. The 
monopoles can be viewed as puncturing holes in the
world sheet through which the core of the space-time theory can be
visualized. 

On the other hand,
interpreting the two-dimensional model from the space-time
point of view, we see at the core a TGT with a
non-compact gauge group, analogous to that proposed by Horava~\cite{horava}. 
The
task we tackle in Section 4 is that of identifying the `matter' fields
that appear around the core by analogy with
the fields $a,b$ above, whose condensation generates the
space-time metric.
Before addressing these issues, though, we first examine more
closely the supersymmetric world-sheet monopole model in an
11-dimensional space time.

~\\
\section{Critical Defects in 11-Dimensional Superstring}

The supersymmetrization of the above
world-sheet defects may be represented using a
sine-Gordon theory~\cite{ovrut,emndbmonop} 
with local $n=1$ supersymmetry, which has
the following monopole deformation operator:
\begin{equation}
V_m = {\bar \psi} \psi : {\rm cos} [{q_m \over \beta^{1/2}_{n=1}}
(\phi(z) - \phi({\bar z}))]:
\label{susymonopole}
\end{equation}
where the $\psi, {\bar \psi}$ are world-sheet fermions with
conformal dimensions $(1/2,0), (0,1/2)$ respectively, and
$\phi$ is a Liouville field. The effective temperature
$1 / \beta_{n=1}$ is related to the matter central charge by~\cite{ovrut}
\begin{equation}
\beta_{n=1} = { 2 \over \pi ( d - 9 )}
\label{susybeta}
\end{equation}
where we assume that $d > 9$. The corresponding vortex
deformation operator is
\begin{equation}
V_v = {\bar \psi} \psi : {\rm cos} [ 2 \pi q_v \beta_{n=1}^{1/2}
(\phi(z) + \phi({\bar z}))]:
\label{susyvortex}
\end{equation}
where $q_v$ is the vortex charge. Including the conformal dimensions
of the fermion fields, we find that the conformal dimensions of the
vortex and monopole operators are
\begin{equation}
\Delta_v = {1 \over 2} + {1 \over 2} \pi \beta_{n=1} q_v^2 = {1 \over 2} +
{q_v^2 \over (d - 9)}, \;\; \\
\Delta_m = {1 \over 2} + {1 \over 8 \pi \beta_{n=1}} q_m^2 = {1 \over 2} +
{q_m^2 (d-9) \over 16}
\label{dimensions}
\end{equation}
respectively. We see that the supersymmetric vortex deformation 
with minimal charge $|q_v| = 1$ is marginal when 
the matter central charge $d = 11$. Below this
limiting value, the vortex deformation is irrelevant. The quantization
condition imposed by single-valuedness of the partition function
tells us that the minimum allowed charge for a dual monopole defect
is $|q_m| = (d - 9)/4$, which is irrelevant for $14.04 > d > 11$.
We therefore see that $d = 11$ is the critical dimension in
which both the supersymmetric vortex and monopole deformations
are marginal. On the world sheet, this corresponds to a
Berezinskii-Kosterlitz-Thouless transition~\cite{xy}, 
with an unstable plasma phase
of free vortex defects in $11 < d < 14.04 $, whilst monopoles are bound 
for matter central charges  $d > 14.04$. We note
that $d=11$ is the maximal
dimensionality of space in which it is possible to have Lorentz-covariant
local supersymmetric theories, 
whilst if one relaxes the requirement of Lorentz 
covariance a 12-- (or higher--) dimensional 
target space may be allowed~\cite{bars}.

In terms of the `temperature' (\ref{susybeta}),
associated with the central-charge deficit of the matter theory, 
the above pattern of phases may be expressed as follows:

\begin{itemize}

\item
(i) $T < T_{BKT-vortex}$, corresponding to $d < 11$: vortices bound, 
monopoles free,

\item
(ii) $T_{BKT-vortex} < T < T_{BKT-monop}$ corresponding to $11 < d <
14.04$: plasma of vortices and monopoles,

\item
(iii) $T > T_{BKT-monop}$ corresponding to $d > 14.04$: monopoles bound,
vortices free.

\end{itemize}
where the two critical Berezinskii-Kosterltiz-Thouless
temperatures are familiar from the two-dimensional XY model~\cite{xy}. 
In our Liouville picture, these critical temperatures correspond to 
critical values of the central charge $d$, as explained above, 
whereas in critical strings such temperatures correspond to 
critical values of the radius of the compactified dimension~\cite{sathiap}.

We have argued that these defects correspond to $D$ branes,
since correlators involving defects and closed-string
operators have cuts for generic values of $\Delta_{v,m}$.
These cause the theory to become effectively that of an open string.
One may then impose Dirichlet boundary conditions on the
boundaries of the effective world sheet, i.e., along the cuts,
obtaining solitonic $D$-brane configurations which become
massless when $d \rightarrow 11$. The 
world-sheet Berezinskii-Kosterlitz-Thouless
transition when $d = 11$ corresponds to the $D$-brane condensation
that occurs in the strong-coupling limit of $M$ theory.
It is known that the low-energy limit of this critical theory
is provided by 11-dimensional supergravity, which
possesses only 3- and 5-brane solitonic solutions.

\section{Anti-de-Sitter Space Time from $D$-Particle Recoil} 

In this section we discuss how eleven-dimensional anti-de-Sitter space
time AdS$_{11}$
arises from our Liouville approach to $D$-brane recoil~\cite{dbrecoil}. 
The $D$ brane is described as above by a world-sheet defect, whose
interaction with a closed-string state 
is described by a pair of logarithmic deformations~\cite{gurarie},
corresponding to the collective coordinate $y_i$
and velocity $u_i$ of the recoiling $D$ particle~\cite{kmw,lizzi}.
Before the recoil, the world-sheet theory with
defects is conformally invariant. However, the
logarithmic operators are slightly relevant~\cite{kmw}, in a 
world-sheet renormalization group sense, with anomalous dimension 
$\Delta = -\frac{\epsilon ^2}{2}$, where $\epsilon$ is a
regularization parameter specified below.   
Thus, the recoiling $D$ particle is no longer described by a 
conformal theory on the world sheet.
To restore 
conformal invariance, one has to invoke 
Liouville dressing~\cite{ddk}, which
increases the target space-time dimensionality
to $d + 1$. Because of the supercriticality~\cite{aben} of the 
central charge $d$ of the 
stringy $\sigma$ model, which had been critical before 
including recoil effects, the Liouville field 
has Minkowski signature in this approach.
In evaluating the
$\sigma$-model path integral, it is convenient to work 
with a Euclidean time $X^0$ in the $d$-dimensional base
space.\footnote{Note
that the time $X^0$ is therefore distinct from the Liouville
time $t$, which necessarily has Minkowski signature.
In the case of gauge theories, the Euclidean time $X^0$
may be thought of as temperature.}
Thus we obtain an effective curved space-time manifold $F$
in $d+1$ dimensions, with signature $(1,d)$, 
which is described~\cite{kanti} by a metric of the form: 
\begin{equation}
G_{00}=-1 \,,\, G_{ij}=\delta_{ij} \,,\,
G_{0i}=G_{i0}=f_i(y_i,t)=\epsilon (\epsilon y_i + u_i t)\, ,\,\,i,j=1,...,d
\label{yiotametric}
\end{equation}
where the regularization parameter $\epsilon \rightarrow 0^+$ is
related~\cite{kmw} to the world-sheet 
size $L$ via 
\be
\epsilon ^{-2} \sim \eta {\rm ln}(L/a)^2,
\label{epsilon}
\ee
where $\eta =-1$ for a 
Liouville mode $t$ of Minkowski signature,
and $a$ a world-sheet short-distance cutoff. The quantities 
$y_i$ and $u_i$ represent the collective coordinates and velocity of a
$d$-dimensional D(irichlet)-particle. 

In our case, the original theory is conformal for $d=11$, 
so the $F$ manifold has twelve dimensions. 
The $D$ particle is a point-like stringy soliton, with $d=11$
collective coordinates
satisfying Dirichlet boundary conditions on the open world sheet 
that appears in the presence of a defect.
Thus the above Liouville theory describes a
12-dimensional space time with {\it two times}~\cite{bars}
if the basis space is taken to have Minkowski signature. The 
fact that the Liouville $\sigma$-model dilaton is
linear in time reflects the non-covariant nature of the 
background~\cite{aben}. This is consistent
with the Lorentz-non-covariant formalism 
of 12-dimensional superstrings~\cite{bars},
which reflects the need to use null vectors 
to construct the appropriate supersymmetries. 
In our approach, this non-covariant nature is a natural consequence of the 
Liouville dressing, prior to supersymmmetry, but this
remark provides for a smooth supersymmetrization of the results. 

We recall~\cite{kanti} that the components of the 
Ricci tensor for the above 12-dimensional $F$ manifold are: 
\begin{eqnarray}
R_{00}&=& -\frac{1}{(1+\sum_{i=1}^{d} f_i^2)^2}\,
\left(\sum_{i=1}^{d} f_i \frac{\partial f_i}{\partial t}
\right)\,\left[\sum_{j=1}^{d} \frac{\partial f_j}
{\partial y_j} \, \left(1+\sum_{k=1, k\neq j}^{d} f_k^2
\right)\right] \\[3mm]
&+& \frac{1}{(1+\sum_{i=1}^{d} f_i^2)}\,\left[\sum_{i=1}^{d}
\frac{\partial^2 f_i}{\partial y_i \partial t}
\left(1+\sum_{j=1, j\neq i}^{d} f_j^2\right)
\right]\\[5mm]
R_{ii}&=&\frac{1}{(1+\sum_{k=1}^{d} f_k^2)^2}\,
\left\{\,\frac{\partial f_i}{\partial y_i}\,
\left(\sum_{j=1}^{d} f_j\,\frac{\partial f_j}
{\partial t}\right)-(1+\sum_{k=1}^{d} f_k^2)\,
\frac{\partial^2 f_i}{\partial y_i \partial t}\right.\nonumber \\[3mm]
&+& \left. \frac{\partial f_i}{\partial y_i} \left[
\sum_{j=1, j\neq i}^{d}\, \frac{\partial f_j}
{\partial y_j}\,(1+\sum_{k=1, k\neq j}^{d} f_k^2)
\right]\right\} \\[5mm]
R_{0i}&=& \frac{f_i}{(1+\sum_{k=1}^{d} f_k^2)^2}\,
\left\{\frac{\partial f_i}{\partial y_i}\,
\left(\sum_{j=1}^{d} f_j \,\frac{\partial f_j}
{\partial t} \right)-\left(1+\sum_{k=1}^{d} f_k^2\right)\,
\frac{\partial^2 f_i}{\partial y_i \partial t}\right\}\\[5mm]
R_{ij}&=& \frac{1}{(1+\sum_{k=1}^{d} f_k^2)^2}\,
f_i \,f_j\,\frac{\partial f_i}{\partial y_i}\,
\frac{\partial f_j}{\partial y_j} 
\label{Ricci}
\end{eqnarray}
We consider below the asymptotic limit: $t >>0$. 
Moreover, we restrict ourselves to the limit
where the recoil velocity $u_i \rightarrow 0$,
which is encountered if the $D$ particle  
is very heavy, with mass $M \propto 1/g_s$, where 
$g_s \rightarrow 0$ is the dual string 
coupling. 
In such a case the closed string state splits into two open ones
`trapped' on the $D$-particle defect. 
The collective coordinates of the latter exhibit quantum
fluctuations of order~\cite{dbrecoil} 
$\Delta y_i \sim |\epsilon ^2 y_i|$.
{}From the world-sheet point of view~\cite{emnmonop}, 
this case of a very heavy $D$ particle corresponds 
to a strongly-coupled defect, since the coupling $e$ of the world-sheet
defect is related to the dual string coupling $g_s$ by
\be
     e\sqrt{\pi/3} \propto \frac{1}{\sqrt{g_s}} 
\label{duality}
\ee
corresponding to a world-sheet/target-space strong/weak coupling duality. 

In the limit $u_i \rightarrow 0$, (\ref{Ricci}) implies that
the only non-vanishing components of the Ricci tensor are: 
\be
R_{ii} \simeq \frac{\partial f_i}{\partial y_i} \left(\sum_{j=1; j\ne i}^{d}
\frac{\partial f_j}{\partial y_j} [1 + {\cal O}(\epsilon ^4)]\right)
\frac{1}{(1 + \sum_{k=1}^{d}f_k^2)^2} \simeq 
\frac{-(d-1)/|\epsilon|^4}{(\frac{1}{|\epsilon|^4} - \sum_{k=1}^{d=11}|y_i|^2)^2} 
+{\cal O}(\epsilon ^{8}) 
\label{limRicci}
\ee
where we have taken into account (\ref{epsilon}) 
and the Minkowski signature of the Liouville 
mode $t$. Thus, in this limiting case
and for large $t >>0$, the Liouville mode decouples
{}from the residual $d$-dimensional manifold. 
We may write (\ref{limRicci}) as
\be 
   R_{ij}={\cal G}_{ij}R 
\label{newricci}
\ee
where ${\cal G}_{ij}$ is a dimensionless diagonal 
metric, corresponding to the line element:
\be
      ds^2=\frac{|\epsilon|^{-4}\sum_{i=1}^{d} 
dy_i^2}{(\frac{1}{|\epsilon|^4} - \sum_{i=1}^{d}|y_i|^2)^2} 
\label{ball}
\ee
This metric describes the {\it interior} 
of a $d$-dimensional ball, 
which is the Euclideanized version of an 
AdS space time~\cite{witten}. 
One can easily check that the curvature of the Minkowski version
of (\ref{ball}) 
is {\it constant} and {\it negative}: $R = -4d(d-1)|\epsilon| ^4$,
independent of the exact location of the $D$ brane. 

It is interesting that the metric of the space time (\ref{ball}) 
exhibits a coordinate singularity
at $\sum _{i=1}^d|y_i|^2=|\epsilon|^{-4}$, which prevents a naive extension 
of the {\it open} ball $B^d$ to the {\it closed} ball 
${\overline B}^d$, including the boundary sphere $S^{d-1}$. The metric
that extends to ${\overline B}^d$ is provided by a conformal
transformation of (\ref{ball})~\cite{witten}: 
\be
    d {\tilde s}^2={\cal F}^2 ds^2 
\label{conformal}
\ee
Choosing, say, 
${\cal F}=|\epsilon |^{-4} - \sum_{i=1}^d|y_i|^2$
results in $d{\tilde s}^2$ being associated with 
the metric on a sphere $S^{d-1}$ of radius 
$\frac{1}{|\epsilon|^2}$.  In general, ${\cal F}$ may be changed by any conformal
transformation, leading to a conformally invariant 
Euclidean $S^{d-1}$ space as the boundary of an 
AdS$_d$ space time, whose metric is invariant under the Lorentz group 
$SO(1,d)$. 

We close this section by recalling that AdS space time is the only
type
of constant-curvature background, for space time dimensionality $d > 2$,
which is consistent with local supersymmetry~\cite{salam}. This is of
vital importance here, because
$D$ branes are stable {\it only} in superstring theories.

\section{Specification of the Local Theory in the AdS$_{11}$ Bulk} 

The important property 
of AdS space times for the next part of our discussion
is the existence of powerful theorems implying
that, if a classical 
field theory is specified in the boundary of the AdS space, then it has a
{\it unique} extension to the bulk~\cite{witten,lee}. 
These theorems underly the
{\it holographic} nature of field/string theories in AdS
space times,  
in the sense that all the information about the 
bulk AdS theory is {\it encoded} in the
boundary theory~\cite{malda,witten}. 

The question now arises, what is the bulk AdS$_{11}$ theory
underlying $M$ theory? Some clues are provided by the
construction of AdS$_{11}$ given in the previous section.
It is natural to look for a theory that becomes local in
the short-distance limit. This theory should, moreover,
have a local symmetry that arises naturally in such a framework.
The natural candidates are gauge theory and gravity,
combined of course with supersymmetry.
In the case of gauge theory, since a conventional quadratic
kinematic term is irrelevant by simple dimensional
counting in more than four space-time dimensions, the most
likely possibility is a Topological Gauge Theory (TGT) of
Chern-Simons type.
Furthermore, since it is known in principle how 
supergravity may arise from an underlying TGT~\cite{chamseddine},
this may also provide a framework in which the notion of a
space-time metric emerges dynamically. The remaining issue is
the choice of gauge supergroup, and this is where the above
construction of the AdS$_{11}$ target space-time geometry 
can provide key input into the short-distance
formulation of $M$ theory as a supersymmetric TGT. 

The minimal supergroup that incorporates the
space-time symmetry of AdS$_{11}$ is $Osp(1|32,R)$, so this
should be a subsupergroup of the conjectured gauge supergroup.
However,
our world-sheet approach provides a hint that the 
full gauge supergroup should be larger than this. We recall
that the departure from criticality inherent to 
the non-trivial $D$-brane recoil that generates AdS$_{11}$
could be absorbed by introducing a time-like Liouville field.
This leads to an underlying $(2,10)$ space-time signature, or
$(1,11)$ in the Euclideanized version needed for an adequate
definition of the path integral. We also recall that
there is an isomorphism between the minimal
supergroup extension of the Lorentz group $SO(1,11)$ and $Osp(1|32,R)
\otimes Osp(1|32,R)$~\cite{holten}. It is therefore natural
to propose that this may play a r\^ole in the supergroup of
the local TGT. Note, however, that the Liouville field
decouples in the conformal limit of zero recoil, and that 
the background field is linear, so any
underlying $SO(1,11)$ or $Osp(1|32,R)\otimes Osp(1|32,R)$
symmetry must be at least spontaneously broken.
The natural minimal possibility is a $Osp(1|32,R)\otimes Osp(1|32,R)
\rightarrow Osp(1|32,R)$ symmetry-breaking pattern, with the
breaking accompanied by the appearance of an AdS$_{11}$ metric
at the world-sheet Berezinskii-Kosterlitz-Thouless transition point.

This is similar to the symmetry-breaking pattern
prposed by~\cite{horava}. 
In that analysis, 
the particular gauge supergroup $Osp(1|32,R)\otimes Osp(1|32,R)$
arose as the minimal supersymmmetric
extension of $Osp(1|32,R)$
with 64 supercharges~\cite{holten,sierra}, 
which are necessary in $M$ theory 
to ensure {\it parity invariance} and
in order to obtain a consistent compactification of $M$ theory 
to a heterotic string with the gauge group $E_8 \otimes
E_8$~\cite{hw}. 
In our Liouville approach, such a group arises 
independently and naturally 
from the presence of the `auxiliary' Liouville field.

\section{Relation to Two-Dimensional Structures}

It is known from the analysis in~\cite{guna2} 
that the contraction of $Osp(1|32, R) \otimes Osp(1|32, R)$ 
with the Poincare symmetry in 11-dimensional space time leads to 
a single diagonal $Osp(1|32, R)$~\footnote{The two factors 
correspond to different spinor representations of the $SO(1,11)$ 
algebra~\cite{guna,guna2}.}. 

Following~\cite{guna}, we now recall that 
$Osp(1|32,R)$ has a two-dimensional maximal subsupergroup 
$Osp(16/2,R)$, which has been argued to capture the dynamics 
of $D0$ particles in the matrix-model approach to M-theory~\cite{matrix}.
The maximal even subgroup 
of $Osp(16/2,R)$ is in turn $Sp(2,R) \otimes SO(16)$. 
It was argued in~\cite{guna} that the 
factor $Sp(2,R)$ corresponds to an AdS$_{2}$
extension of the Poincare group
in the longitudinal directions of the matrix $D$-brane theory. The
connection to this formulation of $D0$ particles supports the
motivation for the $D$-brane interpretation of the world-sheet
defects and the recoil calculation presented earlier.

The singleton representations of $Sp(2,R)$, which live 
on the boundary of AdS$_2$, when   
expanded in a `particle basis',  
consist  of an infinite tower of discrete-momentum states
with ever-increasing quantized $U(1)$ eigenvalues~\cite{sierra}.
Such an infinite tower of states was identified in~\cite{guna} 
with the infinite tower of $D0$ branes with quantized 
longitudinal momentum that appear in matrix theory in the 
infinite-momentum frame~\cite{matrix}. 
This is consistent with the conjecture~\cite{malda}  
that the conformally invariant field theory of the singleton 
representation on the boundary of AdS$_2$ 
is associated with an $N=16$~~~$U(n \rightarrow \infty)$ 
Yang-Mills quantum-mechanical theory~\cite{guna}, which in turn describes 
matrix theory in the infinite-momentum frame. 

We observe here that 
$Sp(2,R)$ is isomorphic to $SO(2,1)$, as well as to AdS$_2$.
This may be related to the possibility of associating 
two-dimensional space times with three-dimensional 
Chern-Simons theories, whose dimensional reduction
leads to AdS$_2$. This possibility recalls 
that of two-dimensional stringy black-hole space times~\cite{emnmonop},
as briefly reviewed in section 2. The appearance of 
a $D$-particle space time 
AdS$_2 \otimes O(16)$ through the `breaking' (described by the
contraction with Poincare symmetry)
of $Osp(1/32,R) \otimes Osp(1/32,R)$ parallels the breaking of
$W_{1+\infty} \otimes W_{1+\infty} \rightarrow
W_{1+\infty}$ in that case. 
In the analysis of~\cite{emnmonop}, 
the association of the two-dimensional model  
with a three-dimensional 
Chern-Simons theory with $CP^1$ `magnon' fields $a,b$ that represent
matter away from the black-hole singularity, 
leads to an interesting symmetry-breaking pattern. 
A renormalization-group analysis has shown that space time appears as a 
non-trivial (infrared) fixed point of the flow.
We present in the next section an alternative formulation of the
appearance of a space-time metric in the full $M$ theory.

The appearance of $D0$ particles provides a nice consistency 
check of the approach we used in section 4, employing 
Liouville $D$-particle recoil to obtain 
AdS$_{11}$ dynamically. We now observe that 
AdS$_2$ structures can be associated with 
topology change in AdS$_{11}$.
To see this, we first review briefly the relevant properties   
of AdS space times~\cite{page,witten}. 
For concreteness, we describe explicitly
the AdS$_4$ case of~\cite{page}. The generalization 
to AdS$_d$, for general $d$, is straightforward~\cite{witten}. 
The Minkowski-signature AdS 
Schwarzschild black hole solution of~\cite{page} 
corresponds to a metric line element of the form: 
\be
   ds^2 = -V (dt)^2 + V^{-1} (dr)^2 + r^2 d\Omega ^2
 \label{adsbh}
\ee
where $d\Omega ^2$ is the line element on a round two-sphere,
$r$ the radial coordinate of the AdS space, 
and $t$ is the time coordinate. 
The Euclidean version of the space time has the topology
$ X_2 =B^2 \otimes S^{n-1}$, where $n=3$ for~\cite{page}.

According to the analysis of~\cite{page}, 
there are two relevant critical temperatures in the AdS black-hole 
system: \\
(i) the specific heat of a gas of 
black holes changes sign at the lowest critical temperature $T_0$. 
For $T < T_0$ there is only radiation, and the topology of 
the finite-temperature space time 
is $X_1 = B^n \otimes S^1$, where $n=3$ in~\cite{page}. \\
(ii) At temperatures above $T_0$,
the topology of the space time 
changes to include black holes, becoming $ X_2 =B^2 \otimes S^{n-1}$. \\
(iv) For temperatures greater than a higher value $T > T_2$,
there is no equilibrium
configuration without a black hole.~\footnote{There is 
also an intermediate temperature $T_1$: $T_0 < T
< T_1 < T_2$, below which
the free energy of the black hole is positive, so the black hole 
tends to evaporate, and above which
the free energy of the configuration with the black hole 
and thermal radiation is lower than the corresponding 
configuration with just thermal radiation,
so that the radiation tunnels to a black-hole state. In our apprach,
this tunnelling may be described by world-sheet instantons~\cite{emn,yung}, 
which we do not
explore further here.}

The extension of the above analysis
to our 11-dimensional AdS$_{11}$ case is straightforward~\cite{witten}.
The important point to notice, for our purpose, is the fact that the 
black-hole space time $X_2$ includes the two-dimensional AdS$_2$ 
space, on the boundary of which live the quantum-mechanical 
$D0$ particles, as discussed in~\cite{guna} and mentioned above. 
Therefore, we associate the group-theoretic observations of~\cite{guna},
with the topology change: $X_1 \rightarrow X_2$ due to 
the appearance of black-hole AdS 
space times. As discussed above, such topology changes may be interpreted
as corresponding to condensation of world-sheet vortex defects. 
Notice, however, that the r\^ole of temperature in this case
is played by the 
central charge deficit of the `matter' theory (\ref{susybeta}).
Thus, when the matter central charge $d$ reaches the critical
value, corresponding to the lowest temperature $T_0$ of~\cite{page}, 
the topology changes, in the sense that the vacuum becomes
dominated by an unstable  plasma of free vortex and monopole 
world-sheet defects. 

The presence of a discrete tower of states living on the boundary 
of AdS$_2$ has been argued~\cite{guna} to be crucial for yielding 
the correct description of the $D0$-particle quantum mechanics. 
Such delocalized states are therefore viewed as 
gravitational degrees of freedom. In this respect, the attention of the 
alert reader should be called to a parallel phenomenon that arises
in two-dimensional stringy black holes with matter.
There, conformal invariance on the world sheet requires,
in a black-hole space time, the coupling to lowest-level zero-momentum
string `tachyon' states of discrete 
solitonic delocalized states that belong to higher string levels~\cite{chaudh}.
In the AdS$_2$ case, the discrete tower of states corresponds to 
a generalization of these lowest zero-momentum `tachyon' states, rather
than the higher-spin delocalized states of the two-dimensional black
hole.

A final topic of relevance here is the tensoring of the
singleton representation in each factor of $Osp(1|32,R)$. 
The resulting theory consists, according to~\cite{guna2}, 
of a doubleton representation that lives on the boundary of 
AdS$_{11}$, which is a ten-dimensional Minkowski space, and is
scale invariant. The next question concerns the coupling of
such scale-invariant theories to superstrings,
which is discussed in the next section. 

\section{Strings as Wilson Loops in TGTs}

Invoking some mean-field conjectures,
Horava~\cite{horava} has shown how one can derive 
the field content of the 11-dimensional supergravity by
calculating Wilson loops $<W(C)>$ of `partons' in his
11-dimensional TGT with gauge supergroup  
$Osp(1|32,R) \otimes Osp(1|32,R)$. Horava~\cite{horava}
did not provide a dynamical model for these `partons' and
Wilson loops. However, our Liouville string approach
provides a natural conjecture for their origins, and, 
as we shall see below, a rather modified proposal for
a local field-theory description of $M$ theory.

As a prelude to our results, we first review briefly the work 
of~\cite{awada}, according to 
which certain scale-invariant 
Green-Schwarz superstring theories in flat target space
are equivalent to Abelian gauge theories.
This equivalence should be understood 
in the sense that
\be 
<W(C)>~\sim~e^{iS_\sigma }
\label{awadaabel}
\ee
where $S_\sigma $ is a world-sheet 
action that encodes the area of the Wilson loop, 
and $W(C)$ is some combination of  
observables, expressed in terms of chiral currents
$e^{\int J.A}$, which 
go beyond the standard Wilson loops. The action $S$ becomes a standard 
world-sheet action if a string scale is generated dynamically
by an appropriate condensation mechanism. 

The analysis of~\cite{awada} was performed 
for four-dimensional target spaces, but it can be 
generalized straightforwardly to six and ten dimensions. 
For reasons of concreteness and calculational simplicity,
we review the analysis in the 
four-dimensional case, where the gauge field theory coupling is 
dimensionless. 

We consider an 
Abelian supersymmetric gauge theory, described by a standard Maxwell 
Lagrangian in a superfield form.~\footnote{The extension of this
analysis to the non-Abelian case raises interesting technical issues
that are currently under study.} 
The connection with string theory emerges by considering the Stokes
theorem on a two-dimensional surface $\Sigma$, 
whose boundary is the loop $C$.
If one parametrizes the curve $C$ by $\lambda$, then one
may write the exponent of the Wilson loop as 
\be
     S_{int} =ie \int _{C} d\tau A( X(\tau)) \frac{\partial }{\partial \tau } X(\tau) 
\label{wilsonexp}
\ee
and the Stokes theorem tells us that
\be
   S_{int} =\frac{ie}{2}\int _{\Sigma (C)} d^2\sigma \epsilon^{ab}
F_{\alpha\beta}, 
\qquad a, b =1,2,
\label{stokes} 
\ee
where $X^M$, $M=1, \dots D$, is a $D$-dimensional 
space-time coordinate for the gauge theory. We denote 
by lower-case Latin indices 
the two-dimensional coordinates of the surface $\Sigma$, 
which plays the r\^ole of the world-sheet of the string 
and is equivalent to the gauge 
theory in question~\cite{polyakov}. The quantity 
$F_{ab} = \partial_a X^M \partial _b X^N F_{NM} =
\partial _a A_b - \partial _b A_a $
is the pull-back of the Maxwell tensor on the world sheet
$\Sigma$, with $A_a$ the corresponding projection of the gauge field 
on $\Sigma$: 
\be
A_a =v_a^M  A_M \qquad ; \qquad v_\alpha ^M \equiv \partial _a X^M , 
a=1,2~; M+1, \dots D(=4).
\label{pullback3}
\ee
{}From a two-dimensional view-point, 
this looks like a Chern-Simons term for a two-dimensional 
gauge theory on $\Sigma$, bounded by the loop $C$. 
The world sheet `magnetic field' corresponding to (\ref{pullback3})
reads:
\be
{\cal B}=\epsilon^{\alpha\beta}\partial_\alpha A_\beta =
\epsilon^{\alpha\beta}\partial_\alpha \partial_\beta X^M A(X)_M
+ \frac{1}{2}\epsilon^{\alpha\beta}\partial_\alpha X^M \partial_\beta X^N
F_{NM}
\label{pullmagn}
\ee
Notice that the presence of world-sheet vortices is associated 
with the first term on the right-hand-side of (\ref{pullmagn}),
whilst world-sheet monopoles are associated with the second term,
which is also gauge invariant in target space.  

The novel observation of~\cite{awada}
is the possibility, in a supersymmetric gauge theory,  
of constructing a second superstring-like 
observable, in addition to the Wilson loop, which is again defined on
the two-surface $\Sigma$, and is consistent with all the symmetries of the
theory.
The second observable is easily understood in the two-dimensional 
superfield formalism:
\be
Z^{{\cal A}} 
\equiv (X^M, \theta ^m, \theta ^{{\dot m}})
\label{sf}
\ee
The pull-back basis $v_a ^M$ in (\ref{pullback3}) is now extended
to $v_a^{{\cal A}} = E^{{\cal A}}_{{\cal B}} \partial _a z^{{\cal B}}$,
with 
the following components~\cite{awada}:  
\bea
&~&v_a^{\alpha{\dot \alpha}} =\partial _z x^{\alpha{\dot \alpha}}
-\frac{i}{2}\left(\theta^\alpha (\sigma) \partial _a \theta ^{{\dot \alpha}} (\sigma)
+ \theta^{{\dot \alpha}} (\sigma) \partial _a \theta ^{\alpha} (\sigma)\right)   
\nn \\
&~& v_a^\alpha = \partial _a \theta ^\alpha (\sigma) \nn \\
&~& v_a^{{\dot \alpha}} = \partial _a \theta ^{{\dot \alpha }} (\sigma)   
\label{superspace}
\eea
in standard notation~\cite{superspace}, where Greek dotted and undotted 
indices denote superspace components, with   $x^{\alpha{\dot \alpha}}
\equiv X^M$, etc..
Following~\cite{awada}, we now define
\bea 
&~& C_{ab}^{\alpha\beta} \equiv \frac{i}{2} 
v_{a{\dot \beta}}^{(\alpha}
v_{b}^{\beta){\dot \beta}},
\nn \\
&~&       C_{ab}^\alpha \equiv v_a^{\alpha{\dot \alpha}}v_{b{\dot
\alpha}}: \nn \\
&~& \qquad 
C^\alpha =\epsilon^{ab}C_{ab}^\alpha , \qquad C^{\alpha\beta}=\epsilon^{ab}C_{ab}^{\alpha\beta}
\label{Cs}
\eea
with similar relations holding for appropriately-defined 
dotted components of $C$, as found in~\cite{awada}. 
Note that $C^{\alpha}$ vanishes in the absence
of supersymmetry, whilst $C^{\alpha\beta}$ exists also
in non-supersymmetric gauge theories.  
In the presence of defects, as we shall later, this is 
no longer the case, and $C^\alpha$ can have non-supersymmetric 
remnants. 

The supersymmetric Wilson loop, expressing the interaction between the 
superparticle and a supersymmetric gauge theory in the approach
of~\cite{awada}, may now expressed as: 
\bea
 &~& W(C)=e^{S_{int}^{(1)}}, \qquad 
S_{int}^{(1)} \equiv \frac{i}{2} \int _{\Sigma (C)} d^2\sigma \epsilon^{ab}
{\cal F}_{ab} \nn \\
&~&{\cal F}_{ab} \equiv \epsilon_{ab}\{\frac{1}{2}C^{\alpha\beta}(\sigma)
D_\alpha W_\beta (x(\sigma),\theta (\sigma)) + \nn \\
&~& C^\alpha (\sigma) W_\alpha
(x(\sigma),\theta (\sigma)) + h.c. \} 
\label{susywilson}
\eea
where $W_\alpha (x(\sigma),\theta (\sigma))$ is the chiral superfield 
of the supersymmetric Abelian gauge theory. 
The exponent $S_{int}^{(1)}$ 
clearly reduces to the standard expression (\ref{wilsonexp}) 
in non-supersymmetric cases.

It was pointed out in~\cite{awada} that 
there is a second superstring-like observable, $\Psi (\Sigma )$, 
defined on the world-sheet surface $\Sigma$,
which is constructed out of the $C_{ab}^\alpha $ components, 
which therefore - in the absence of world-sheet defects - 
exists only in supersymmetric gauge theories:
\be
 \Psi (\Sigma ) \equiv e^{iS_{int}^{(2)}}, \qquad 
S_{int}^{(2)} \equiv \kappa \int _{\Sigma (C)} d^2\sigma 
\sqrt{-\gamma}\gamma ^{ab}C_{ab}^\alpha (\sigma) 
W_\alpha (x(\sigma), \theta (\sigma)) + h.c.
\label{second}
\ee
where $\gamma ^{ab}$ is the metric on $\Sigma$. This term,
unlike the standard Wilson loop, 
is not
a total world-sheet derivative, and therefore lives in the bulk of the 
world-sheet $\Sigma$, and depends on the metric $\gamma $. 
The coupling constant $\kappa$ 
is defined classically as an independent coupling. However,
one expects that quantum effects will relate it to the gauge coupling 
constant $e$, a point we return to later. 
This second observable has been expressed in~\cite{awada} 
in terms of `chiral' currents on
$\Sigma$, located at the string source:
\bea
&~&   S_{int}^{(2)} =\int d^6Z \left({\cal J}^\alpha 
W_\alpha + h.c. \right), \nn \\
&~& {\cal J}^\alpha \equiv \kappa \int _{\Sigma (C)} d^2\sigma \sqrt{-\gamma}\gamma ^{ab}C_{ab}^\alpha (\sigma) \delta ^{(6)} (Z-Z(\sigma)),  
\nn \\
&~& \delta ^{(6)} (Z-Z(\sigma)) =\delta ^{(4)} (Z-Z(\sigma))\left(\theta - \theta (\sigma)\right)^2 
\label{chiral}
\eea
Upon integrating out the gauge field components in (\ref{second}), 
i.e., considering the vacuum expectation value 
$<\Psi (\Sigma )>$, where $< \dots >$ denotes
averaging with respect to the Maxwell action for the gauge field, 
the authors of~\cite{awada} 
have obtained a superstring-like action, which is 
scale invariant in target space, as well as on the world sheet:
\bea
 &~&   <\Psi (\sigma )>_{Maxwell} =e^{S_0 + S_1}  \nn \\
&~& S_0 \equiv \frac{\kappa _0^2}{16\pi} \int _{\Sigma } d^2 \sigma 
\sqrt{-\gamma}\gamma ^{ab}v_a^Mv_b^N\sigma _M\sigma _N, \nn \\ 
&~& S_1 \equiv \frac{\kappa _1^2}{4\pi} \int _{\Sigma (C)} 
\sqrt{-\gamma} \gamma ^{ab}v_a^M v_b^N \eta_{MN} \sigma^K\sigma_K
\label{gaugefield}
\eea
where upper-case Latin indices denote target-space indices ($M,N,K=1,
\dots d(=4$)), and 
\bea
&~& v_a^M \equiv \partial_a X^M(\sigma) - 
i{\overline \theta}^m(\sigma)\Gamma ^M \partial _a \theta _m  (\sigma) 
\nn \\
&~&    \sigma^M=\frac{\sqrt{-\gamma}\epsilon^{ab}}{\sqrt{{\rm det}G}}\partial_a v_b^M, \quad G_{ab} \equiv v_a^Mv_b^N\eta_{MN} 
\label{fourcomp}
\eea
in standard four-component notation in target space, where the $m$ are
spinor indices, and the $\Gamma ^M$ are four-dimensional Dirac matrices. 
The dimensionless coupling constants $\kappa _{0,1}$ appear arbitrary 
at the classical level, but one expects them to be related
to the dimensionless gauge coupling $e$, in the quantum theory.

The important observation in~\cite{awada} was the fact that the 
world-sheet action $S_2$ (\ref{second}) resembles the
classical Green-Schwarz superstring action in flat four-dimensional 
target space, provided that condensation occurs for the composite field
\be
\Phi \equiv \sigma^M \sigma_M,\qquad M=1, \dots D(=4), 
\label{composite}
\ee
which may be interpreted as the dilaton in target space, so that 
\be
\frac{\kappa _1^2}{4\pi}<\Phi>=\mu_{string~tension}
\label{tension}
\ee
Such condensation would enable the string tension $\mu$ to be obtained
from a gauge theory without dimensionful parameters.
The question of physical interest
is what causes this condensation, which presumably 
is responsible for a `spontaneous breaking' of 
the scale invariance of four-dimensional string theory. 

We note that the above definition of
the composite field $\Phi$
may be extended to higher-dimensional 
cases, which are of interest to us in this $M$-theory application, 
by simply contracting the integrand of (\ref{second}) 
with appropriate powers of $\Phi $ so as to make the coupling constant 
$\kappa$ dimensionless. This can be understood as the definition of the 
superstring observable $\Psi $ in more than four target space-time 
dimensions. It is important 
to note that the dimensionality of the coupling constant $\kappa$
is the same as that of the gauge theory coupling $e$ only 
for space-time dimensionality four, six and ten, where supersymmetric
theories exist as well. 

We now present a scenario for the formation of 
such a condensate, which is an extension of the ideas of~\cite{witten}. 
We associate the second observable (\ref{second}) 
of~\cite{awada} to the condensation 
of non-trivial world-sheet defects. 
The scenario is based on the recent 
interpretation~\cite{witten} of confinement in purely gluonic
non-Abelian gauge theories, by means of a 
holographic principle encoding 
confinement quantum physics in the classical 
geometry of uncompactified AdS$_5$ space times. 
In such a picture, four-dimensional conformally-invariant 
Minkowski space time is 
viewed as the boundary of an AdS$_5$
space time, in the spirit of~\cite{malda,witten}. 
Macroscopic black holes   
disappear at temperatures below $T_0$,
where only radiation-dominated
universes exist~\cite{page}, as discussed above. 
Witten~\cite{witten}, has associated 
this temperature to a confining-deconfining phase 
transition for quarks, and stressed the fact that,
for spatial Wilson loops the area law and the associated string tension
are obtained only for finite-temperature field theory, because of the
conformal invariance of the zero-temperature four-dimensional gauge theory, 
whose renormalization-group $\beta$ function vanishes identically. 

It is important to notice that condensation of 
composite operators  
can occur for both vortex (\ref{solution}) 
and monopole (\ref{solution2}) configurations.
Let us first examine the case 
of the manifold with topology $X_1 = B^n \otimes S^1$.  
In this case, one may consider the role of vortices on the world sheet 
of the string, wrapped around the compact dimension $S^1$. The
condensation of such vortices, bound into pairs with antivortices, 
results in the quantity $\Phi $  (\ref{composite})
acquiring a non-trivial vacuum expectation value $<\Phi > \ne 0$.
In the classical AdS picture described above, 
it is the non-trivial Planckian dynamics 
of a five-dimensional space time which is responsible for the above 
phenomenon. In such a case the standard Berezinskii-Kosterlitz-Thouless
phase transition temperature for vortex condensation 
on the world 
sheet of critical strings~\cite{sathiap}
may be identified with $T_0$ of the AdS Black Hole space time.

\section{A Conjecture for the Structure and Dynamics of $M$ Theory} 

We are now equipped to formulate a conjecture
on the structure and dynamics of  $M$ theory. 
Since ten-dimensional Minkowski space time
may be considered as the boundary of AdS$_{11}$, which arises
dynamically through the Liouville 
dressing of a quantum-fluctuationg
$D$ particle, we invoke the analysis of~\cite{awada} for Abelian 
gauge theories, to conjecture that there exists a
conformal field theory on the ten-dimensional Minkowski
space time ${\cal M}$, which is dual
to AdS$_{11}$ in the following sense~\cite{malda}:
\be 
   <e^{\int_{{\cal M}} \phi_0 {\cal O}} >_{CFT} =Z_{AdS_{11}}(\phi) 
\label{doubleton}
\ee
where the ${\cal O}$ are appropriate local operators of the conformal 
field theory. This is supported by the fact that the conformal group 
of ${\cal M}$ is the same as $SO(2,10)$. Gunaydin has suggested~\cite{guna2} 
that the
conformal field theory might be the doubleton field theory living on
${\cal M}$. 

At this point, we should also remark 
that, according to Horava~\cite{horava}, the correspondence with the 
11-dimensional supergravity occurs through partons of the gauge group 
$Osp(1/32, R)\otimes Osp(1/32,R)$ in an 11-dimensional Chern Simons 
topological Gauge theory. This is different from 
the conjecture (\ref{doubleton}), which seems more natural from the 
point of view of~\cite{awada,malda}. 
However, the approach of~\cite{horava} may be connected to the conjecture
(\ref{doubleton}), if one makes a connection between the 
Chern-Simons gauge theory on $Osp(1/32,R)\otimes Osp(1/32,R)$
used in~\cite{horava} with an appropriate string theory
in a gravitational background in $12$ space-time dimensions.
Such a correspondence should be expected from the isomorphism of 
the supergroup extension of
$SO(1,11)$ with $Osp(1/32,R)\otimes Osp(1/32,R)$. 
Summing this string theory, of Liouville type, 
over world-sheet genera in gravitational 12-dimensional backgrounds,
as discussed above, produces a path integral over string 
backgrounds, as discussed in~\cite{emndbraneliouville}. 

The fact that the 12-dimensional space-time backgrounds are not
Lorentz covariant is not a problem, in view of the non-covariant 
form of the Liouville theory. 
The basic formula of~\cite{emndbraneliouville} for such a string-theory
space quantization may be summarized as:
\be
   {\cal Z}=\int Dg^i e^{-C[g^j] + 
\int d^2\sigma \partial_\alpha X^M \partial^\alpha X^N < \partial^\gamma J_{\gamma,M} \partial ^\delta J_{\delta,N}>_{g}} + \dots 
\label{stringtheoryspace}
\ee
where the $\{ g^i \}$ denotes an appropriate set of backgrounds, 
including the graviton, the $\dots $ denote antisymmetric tensor and other 
string sources,  
$C[g]$ is the Zamolodchikov central-charge action, which plays the 
r\^ole of an effective (low-energy) target-space action of the string
theory, 
and the $J_M$ are Noether currents, corresponding to the 
translation invariance in target space time of the string. 
The measure of integration $Dg^i$ arises from 
summing over world-sheet genera. 

It was argued in~\cite{emndbraneliouville} that condensation 
in the two-point function of the currents may lead to 
well-defined normalizable metric backgrounds, 
coupled to the string source,
provided that the currents are logarithmic~\cite{gurarie}.  
The same would be true for 
antisymmetric tensor source terms, etc..
This means that there are appropriate 
$p$-brane solutions, obtainable as saddle points, with respect to 
the various backgrounds ${\bar g}^i$, of the  
partition function
(\ref{stringtheoryspace}). For instance, for graviton backgrounds,
\be
    \frac{\delta}{\delta G_{MN}} C[{\bar g}] + V_{MN} =0
\label{saddle}
\ee
where the graviton vertex operator $V_{MN}$ plays the 
r\^ole of an external string source. 
  
Among such solutions there would be AdS backgrounds, 
such as the ones discussed above, which are already known to be
associated with logarithmic recoil operators~\cite{kmw}. 
Based on these considerations, one may expect a connection of
the quantum theory of~\cite{horava}:
$< W(C)> \sim {\cal Z}$, 
with the classical AdS
theory, c.f. the mean-field approximation in (\ref{saddle}).
In that case, a classical AdS effective action
which has a holographic property~\cite{witten,malda}
may be viewed
as the mean-field result of the quantum theory of the 
fluctuating stringy 
background (\ref{stringtheoryspace}). In this picture, 
the appearance of a 
doubleton conformal field theory (\ref{doubleton}) 
on the boundary Minkowski space time
may well be valid~\footnote{In the sense that, at least at present, 
we do not see an apparent conflict between the two conjectures.},
thereby unifying the conjecture (\ref{doubleton}), also made
in~\cite{guna2},
with that made in \cite{horava}.  
However, we re-emphasize that the above discussion should be considered
conjectural at the present stage.
~\\
\noindent
{\bf Acknowledgements}
~\\
The work of D.V.N. is supported in part by the Department of Energy
under grant number DE-FG03-95-ER-40917.


\begin{thebibliography}{99} 

\bibitem{dbranes} J. Polchinski, Phys. Rev. Lett. 75 (1995), 184;
\par C. Bachas, Phys. Lett. B374 (1996), 37;
\par J. Polchinski, S. Chaudhuri and C. Johnson, heo-th/9602052
and references therein; 
\par J. Polchinski, TASI lectures on $D$ branes,hep-th/9611050, and references therein; 
\par E. Witten, Nucl. Phys. B460 (1996), 335; 
\par For a recent review see: A. Sen, hep-th/9802051, and references therein.

\bibitem{duff} For a review see: M. Duff, Int. J. Mod. Phys. A11 (1996), 
5623, and references therein.

\bibitem{matrix} T. Banks, W. Fischler, S. Shenker and L. Susskind, 
Phys. Rev. D55 (1997), 5112. 


\bibitem{emndbmonop} J. Ellis, N.E. Mavromatos and D.V. Nanopoulos,
Mod. Phys. Lett. A12 (1997), 2813. 



\bibitem{aben} I. Antoniadis, C. Bachas, J. Ellis  and D.V. Nanopoulos, 
Phys. Lett. B211 (1988), 383; Nucl. Phys. B328 (1989), 117. 



\bibitem{ddk} F. David, Mod. Phys. Lett. A3 (1988), 1651;
\par J. Distler and H. Kawai, Nucl. Phys. B321 (1989), 509;
\par see also: N.E. Mavromatos and J. L. Miramontes, Mod. Phys. Lett. 
A4 (1989), 1847. 


\bibitem{emn} J. Ellis, N.E. Mavromatos
and D.V. Nanopoulos, Phys. Lett. B293 (1992), 37;
\par Mod. Phys. Lett. A10 (1995), 425;
\par Lectures presented at the
{\it Erice Summer School, 31st Course: From Supersymmetry to the
Origin of Space-Time},
Ettore Majorana Centre, Erice, July 4-12
1993 ; hep-th/9403133, `Subnuclear Series' Vol. 31, 
(World Scientific, Singapore 1994), p.1. 


  
\bibitem{wittenbh} E. Witten, Phys. Rev. D44 (1991), 314. 



\bibitem{emnmonop} J. Ellis, N.E. Mavromatos and D.V. Nanopoulos, 
Phys. Lett. B289 (1992), 25. 



\bibitem{horava} P. Horava, hep-th/9712130. 


\bibitem{eguchi} T. Eguchi, Mod. Phys. Lett. A7 (1992), 85. 


\bibitem{emnorigin} J. Ellis, N.E. Mavromatos and D.V. Nanopoulos, 
Phys. Lett. B288 (1992), 23. 



\bibitem{holten} J. W. Van Holten and A. Van Proyen, 
J. Phys. A15 (1982), 3763. 




\bibitem{singl} C. Fronsdal, Phys. Rev. D12 (1975), 3819;
\par M. Flato and C. Fronsdal, Lett. Math. Phys. 2 (1978), 421. 



\bibitem{guna} M. G\"unaydin and D. Minic, hep-th/9802047 
and references therein. 


\bibitem{guna2} M. G\"unaydin, hep-th/9803138. 



\bibitem{awada} M. Awada and F. Mansouri, 
Phys. Lett B384 (1996), 111; {\it ibid.} B387 (1996), 75.





\bibitem{sathiap} I. Kogan, JETP Lett. (1987), 709;
\par B. Sathiapalan, Phys. Rev. D35 (1987), 709;
\par J. Atick and E. Witten, Nucl. Phys. B310 (1988), 291;
\par A.A. Abrikosov Jr. and I. Kogan, Int. J. Mod. Phys. A6 (1991), 1501. 



\bibitem{ovrut} B. Ovrut and S. Thomas, Phys. Lett. B237 (1991), 292. 



\bibitem{xy} V.L. Berezinskii, JETP 34 (1972), 610 ;
\par J.M. Kosterlitz and D.J. Thouless, J. Phys. C6 (1973), 1181. 



\bibitem{bars} I. Bars and C. Kounnas, Phys. Rev. D56 (1997), 3664;
\par I. Bars and C. Deliduman, Phys. Rev. D56 (1997), 6579;
\par H. Nishino, hep-th/9703214. 


\bibitem{dbrecoil} J. Ellis, N.E. Mavromatos and D.V. Nanopoulos, 
hep-th/9609238, Int. J. Mod. Phys. A, in press; and
Mod. Phys. Lett. A12 (1997), 1759.




\bibitem{gurarie} V. Gurarie, Nucl. Phys. B410 (1993); 
\par For applications relevant to our context, see:
\par A. Bilal and I. Kogan, Nucl. Phys. B449 (1995), 569;
\par I. Kogan and N.E. Mavromatos, Phys. Lett. B375 (1996), 11;
\par J.S. Caux, I. Kogan and A.M. Tsvelik, Nucl. Phys. B466 (1996), 444.  
\par N.E. Mavromatos and R.J. Szabo, hep-th/9803092. 


\bibitem{kmw} I. Kogan, N.E. Mavromatos and J.F. Wheater, Phys. Lett. B387 (1996), 483. 



\bibitem{lizzi} F. Lizzi and N.E. Mavromatos, Phys. Rev. D55 (1997), 7859. 



\bibitem{kanti} J. Ellis, P. Kanti, N.E. Mavromatos, D.V. Nanopoulos 
and E. Winstanley, Mod. Phys. A13 (1998), 303. 








\bibitem{witten} E. Witten, hep-th/9802150; hep-th/9803131. 




\bibitem{salam} A. Salam and E. Sezgin, 
{\it Supergravity in Different Dimensions} (World Scientific, 1989). 




\bibitem{lee} R. Graham and J. Lee, Adv. Math. 87 (1991), 186. 


\bibitem{malda} J. Maldacena, hep-th/9711200. 


\bibitem{chamseddine} A. Chamseddine, Phys. Lett. B233 (1989), 291. 


\bibitem{sierra} M. G\"unaydin, G. Sierra and P.K. Townsend, Nucl. Phys. 
B274 (1986), 429. 


\bibitem{hw} P. Horava, and E. Witten, Nucl. Phys. B460 (1996), 506; 
\par E. Witten, hep-th/9609122. 



\bibitem{page} S. Hawking and D. Page, Com. Math. Phys. 87 (1983), 577. 


\bibitem{yung} A.V. Yung, Int.J.Mod. Phys. A9 (1994), 591;
{\it ibid.} A10 (1995), 1553.  




\bibitem{chaudh} S. Chaudhuri and J. Lykken, Nucl. Phys. B396 (1993), 270. 




\bibitem{polyakov} A. M. Polyakov, Nucl. Phys. B72 (1974), 461; 
{\it ibid} B268 (1986), 406;
{\it Gauge Fields and Strings} (Harwood, 1987);
\par see also: A.M. Polyakov, hep-th/9711002;
\par S. Gubser, I. Klebanov and A.M. Polyakov, hep-th/9802109.




\bibitem{superspace} S. J. Gates {\it et al.}, 
{\it Superspace: 1001 Lessons in Supersymmetry} 
(Benjamin/Cummings, 1983).  




\bibitem{emndbraneliouville} J. Ellis, N.E. Mavromatos and D.V. Nanopoulos,
Int. J. Mod. Phys. A12 (1997), 2639.



\end{thebibliography}
\end{document}